\documentclass[conference]{IEEEtran}
\IEEEoverridecommandlockouts
\usepackage{cite}
\usepackage{amsmath,amssymb,amsfonts}
\usepackage{algorithmic}
\usepackage{graphicx}
\usepackage{textcomp}
\usepackage{float}
\usepackage{xcolor}
\usepackage{enumitem}
\usepackage[ruled,vlined,linesnumbered]{algorithm2e}
\usepackage{booktabs}
\usepackage{makecell}
\usepackage{graphicx}
\usepackage[inkscapelatex=false]{svg}
\usepackage{subcaption}
\usepackage{hyperref}
\def\BibTeX{{\rm B\kern-.05em{\sc i\kern-.025em b}\kern-.08em
    T\kern-.1667em\lower.7ex\hbox{E}\kern-.125emX}}
\begin{document}

% --- ADD THIS FROM YOUR DISCOVERY ---
\makeatletter 
\newcommand{\linebreakand}{% 
\end{@IEEEauthorhalign} 
\hfill\mbox{}\par 
\mbox{}\hfill\begin{@IEEEauthorhalign} 
}
\begingroup
\newcommand{\blankfootnote}[1]{%
  \begingroup
  \renewcommand{\thefootnote}{}%
  \footnote{#1}%
  \addtocounter{footnote}{-1}%
  \endgroup
}
\makeatother
% ------------------------------------

\title{LLM-PRISM: Characterizing Silent Data Corruption from Permanent GPU Faults in LLM Training}

\author{
\IEEEauthorblockN{Abhishek Tyagi}
\IEEEauthorblockA{\textit{University of Rochester}\\
Rochester, NY, USA \\
atyagi2@cs.rochester.edu}
\and
\IEEEauthorblockN{Saurabh Hukerikar}
\IEEEauthorblockA{\textit{NVIDIA Corporation}\\
Santa Clara, CA, USA \\
shukerikar@nvidia.com}
\and
\IEEEauthorblockN{Nirmal Saxena}
\IEEEauthorblockA{\textit{NVIDIA Corporation}\\
Santa Clara, CA, USA \\
nsaxena@nvidia.com}
\and
\IEEEauthorblockN{Yanxiang Huang}
\IEEEauthorblockA{\textit{NVIDIA Corporation}\\
Santa Clara, CA, USA \\
yanxiangh@nvidia.com}
\linebreakand
\IEEEauthorblockN{Philip Shirvani}
\IEEEauthorblockA{\textit{NVIDIA Corporation}\\
Santa Clara, CA, USA \\
pshirvani@nvidia.com}

 \and% <--- YOUR CUSTOM COMMAND FORCES THE 5-2 CENTERED BREAK

\IEEEauthorblockN{Chung-Hsuan Tung}
\IEEEauthorblockA{\textit{Duke University}\\
Durham, NC, USA \\
ct297@duke.edu}
\and
\IEEEauthorblockN{Yuhao Zhu}
\IEEEauthorblockA{\textit{University of Rochester}\\
Rochester, NY, USA \\
yzhu@rochester.edu}

%\thanks{\textsuperscript{*}Work done while the author was an intern at NVIDIA.}
}

%!TEX root=paper.tex
% Put all your macros here
\newcommand{\website}[1]{{\tt #1}}
\newcommand{\program}[1]{{\tt #1}}
\newcommand{\benchmark}[1]{{\it #1}}
\newcommand{\fixme}[1]{{\textcolor{red}{\textit{#1}}}}
\newcommand{\takeaway}[1]{{\textcolor{teal}{Takeaway:\textit{#1}}}}

\newcommand*\circled[2]{\tikz[baseline=(char.base)]{
            \node[shape=circle,fill=black,inner sep=1pt] (char) {\textcolor{#1}{{\footnotesize #2}}};}}

\ifx\figurename\undefined \def\figurename{Figure}\fi
\renewcommand{\figurename}{Fig.}
\renewcommand{\paragraph}[1]{\textbf{#1} }
\newcommand{\figline}{{\vspace*{.05in}\hline}}

\newcommand{\cmark}{\ding{51}}%
\newcommand{\xmark}{\ding{55}}%
\newcommand{\Sect}[1]{Sec.~\ref{#1}}
\newcommand{\Fig}[1]{Fig.~\ref{#1}}
\newcommand{\Tbl}[1]{Tbl.~\ref{#1}}
\newcommand{\Eqn}[1]{Eqn.~\ref{#1}}
\newcommand{\Apx}[1]{Apdx.~\ref{#1}}
\newcommand{\Alg}[1]{Algo.~\ref{#1}}
\newcommand{\CifarComb}{CIFAR10 and CIFAR100~\cite{krizhevsky2009learning} }
\newcommand{\Cifarten}{CIFAR10 }
\newcommand{\Cifarhun}{CIFAR100 }
\newcommand{\ImgNet}{ImageNet-1K~\cite{deng2009imagenet}}
\newcommand{\method}{DynaDiag\xspace}
\newcommand{\methodperm}{PA-DST\xspace}
\newcommand{\TopK}{\(\mathrm{TopK}\) }
\newcommand{\smat}{SmaT }

\newcommand{\specialcell}[2][c]{\begin{tabular}[#1]{@{}c@{}}#2\end{tabular}}
\newcommand{\note}[1]{\textcolor{red}{#1}}

\newcommand{\proj}{\textsc{Cicero}\xspace}
\newcommand{\algo}{\textsc{SpaRW}\xspace}
\newcommand{\mode}[1]{\underline{\textsc{#1}}\xspace}
\newcommand{\sys}[1]{\underline{\textsc{#1}}}

\newcommand{\no}[1]{#1}
\renewcommand{\no}[1]{}
\newcommand{\RNum}[1]{\uppercase\expandafter{\romannumeral #1\relax}}

\def\cG{{\mathcal{G}}}
\def\cF{{\mathcal{F}}}
\def\cI{{\mathcal{I}}}
\def\cN{{\mathcal{N}}}
\def\bh{{\mathbf{h}}}
\def\bp{{\mathbf{p}}}

% checkmark and xmark in the pifont package
%\newcommand{\cmark}{\ding{51}}
%\newcommand{\xmark}{\ding{55}}

\maketitle
\begin{abstract}
Large-scale LLM training is increasingly susceptible to hardware defects stemming from manufacturing escapes and silicon aging. These defects manifest as Silent Data Corruption (SDC) that perturb gradients and parameters throughout the training process. We present LLM-PRISM\footnote{PRISM stands for Permanent and Intermittent SDC Modeling}, a methodology to characterize LLM pre-training resilience to hardware faults. LLM-PRISM couples RTL-level GPU fault simulation with a stochastic injection engine embedded in Megatron-LM. Through 7,664 training runs across FP16, BF16, and FP8 regimes, we analyze how fault type, rate, and numeric format govern resilience. We find that while LLMs resist low-frequency faults, impact is highly non-uniform; critical datapaths and specific precision formats can induce catastrophic divergence even at moderate fault rates. This study provides the first hardware-grounded, pre-training characterization of SDC resilience.

\end{abstract}


%\blankfootnote{PRISM stands for Permanent and Intermittent SDC Modeling}

\section{Introduction}
\label{sect:intro}
The rapid growth of large language models (LLMs) has driven the deployment of massive-scale AI infrastructure, with state-of-the-art training runs now executing on clusters with tens of thousands GPUs spanning weeks to months~\cite{grattafiori2024llama}. As these systems scale, the reliability of the underlying hardware becomes increasingly crucial~\cite{Saxena:Micro:2026}. Silent data corruption (SDC), faults that lead to incorrect computations without triggering detectable errors ~\cite{dixit2021silent, hochschild2021cores, wang2023understanding}, poses a serious threat to the correctness, stability, and reproducibility of LLM training~\cite{ocp2026sdcinai}. Even minor deviations during training can compound through billions of updates, potentially degrading convergence behavior, pre-training accuracy, or downstream model accuracy and/or performance in subtle and costly ways.

Most prior studies assessing the resilience of deep learning workloads have focused on transient faults \cite{sun2025demystifying,yu2025exploring,tyagi2024characterizing,tyagi2022thales}. These non-persistent faults, often caused by cosmic radiation or particle strikes manifest as bit-flips that momentarily corrupt the architectural state of processors. 
%without inducing permanent hardware degradation. 
However these analyses overlook another class of reliability challenges — permanent faults, which can manifest either continuously or intermittently depending on operating conditions.
Permanent faults stem from manufacturing defects that escape testing, latent defects that become active during the early operational lifetime, or silicon wear-out mechanisms \cite{mitra2025escapes}. 
%such as electromigration (EM)~\cite{black2005electromigration}, bias temperature instability (BTI)~\cite{grasser2011paradigm}, and time-dependent dielectric breakdown (TDDB)~\cite{degraeve1998tddb}. 
Unlike transient faults, these errors can silently corrupt computations over long periods or intermittently manifest depending on workload activity, temperature, or voltage conditions. As semiconductor technology advances towards sub-5nm process nodes and with the growing architectural complexity of modern AI accelerators featuring billions of transistors, extensive on-chip interconnects, and heterogeneous compute fabrics, this class of faults is becoming increasingly prevalent. 

Emerging evidence from LLM production training workloads indicates that permanent or marginal hardware defects that causes SDC introduce subtle numerical errors that do not trigger failures but quietly perturb gradients and parameters~\cite{elsen2023errant, metallama3_2024, gemini2023gemini}. However their cumulative effect on training dynamics and final model behavior remains largely unexplored and not well understood. Therefore, the core focus of this work is on characterizing the sensitivity of LLM training to silent data corruption caused by permanent faults, with particular emphasis on their intermittent manifestations. 
%\fixme{what is an intermittent fault? is it a kind of permanent fault? this is brought up abruptly.}. 

\paragraph{This work makes the following key contributions:}  
\begin{itemize}
    \item We develop a fault model, a quantitative abstraction of how underlying hardware defects perturb architectural state, for permanent hardware defects, including their intermittent manifestations, relevant to modern GPUs, and demonstrate how it can be instantiated within NVIDIA’s Megatron-LM~\cite{shoeybi2019megatron} framework to enable systematic fault injection in large-scale LLM training.
%    \fixme{what's a fault model anyway?}
    \item We study how such permanent faults perturb the LLM training pipeline, analyzing their impact on loss trajectories, convergence behavior, and final model quality. 
    \item We perform an end-to-end evaluation of LLM training simulating thousands of permanent faults across multiple precision types (FP16, BF16, FP8). 
    \item We evaluate widely deployed runtime monitoring mechanism like NaN/Inf detectors and quantify its effectiveness and blind spots for detecting SDC induced by permanent and intermittent faults.
\end{itemize}
Understanding how these faults propagate through the training pipeline can inform fault-aware system design, more robust algorithmic methods to detect and correct errors, or the use of selective redundancy strategies. 

\paragraph{Paper Overview:}Section~\ref{sec:background} reviews permanent hardware fault mechanisms and their implications for modern LLM training. Section~\ref{sec:method} presents LLM-PRISM, our methodology for characterizing SDC resilience in LLM pre-training.  Section~\ref{sec:exp} describes the experimental setup. Section~\ref{sec:results} presents the main findings from a large-scale empirical study, spanning RTL-level fault characterization and end-to-end effects on training dynamics, including the behavior of NaN/Inf detection across precision formats. Section~\ref{sec:limit} discusses limitations, and Section~\ref{sec:conclusion} concludes with implications for fault-aware LLM training systems.
\section{Background}
\label{sec:background}

\subsection{Permanent Faults}
\label{sec:back:perm}

Permanent faults arise from three distinct root causes: manufacturing test escapes, where defective devices pass production screening and only reveal themselves under field conditions; latent early-life defects, which surface during the initial operational period as marginal structures fail under stress; and aging-induced failures, which develop progressively over a chip’s lifetime as wear-out mechanisms such as BTI~\cite{grasser2011paradigm}, Hot Carrier Injection (HCI)~\cite{hu1985hot}, TDDB~\cite{degraeve1998tddb}, and Electromigration (EM)~\cite{black2005electromigration} degrade device characteristics.
At the logical level, permanent faults may manifest as stuck-at faults in combinational logic, flip-flops, or RAM arrays, where a node is permanently fixed at a logic '0' or '1' regardless of the applied input — these are \textbf{timing-independent faults} whose effect is deterministic and persistent across all operating conditions. Partial degradation of transistors or interconnects introduces small delay defects that reduce the available timing margin in a circuit path. When these delays grow sufficiently large, they cause violations of setup or hold time constraints, leading to incorrect values being latched or propagated. Such faults are classified as \textbf{timing-dependent permanent faults} or \textit{intermittent faults} that exhibit marginality - they tend to manifest selectively at specific voltage or frequency operating points. Such timing-dependent, intermittent faults are a critical concern in advanced process nodes and in compute elements and RAM arrays — where aggressive voltage/frequency scaling and current densities exacerbate marginal timing behavior. %In this work, a fault model is adopted that captures the behavioral signatures of both timing-independent and timing-dependent permanent faults rooted in these physical degradation mechanisms.

\subsection{Silent Data Corruption (SDC)}
\label{sec:sec:sdc}
%\fixme{This section contains the background on what SDCs are and how they can be caused by different hardware faults such as transient faults, intermittent faults and permanent faults.}
SDC denotes a class of hardware errors in which a system produces incorrect computational results without triggering explicit failure signals or program crashes. These errors typically manifest as bit flips in memory or logic, unintentionally altering the binary representations of parameters, activations, or gradients during LLM training. Unlike fail-stop errors that halt execution, SDC allows training to proceed with a corrupted state, perturbing intermediate computations and silently altering the optimization trajectory. The root causes of SDCs are diverse, spanning transient, intermittent, and permanent hardware faults that degrade execution reliability, with intermittent and permanent faults often arising from latent silicon defects, circuit aging, or aggressive voltage and frequency scaling that create “unhealthy nodes” which consistently or sporadically produce errors under high-stress conditions. In the context of LLM training, such silent perturbations can accumulate over billions of update steps, making SDC particularly problematic because it undermines both numerical correctness and the reproducibility of large-scale training runs.
%Transient faults, often referred to as soft errors, are typically induced by environmental factors such as high-energy particle strikes (e.g., alpha particles or cosmic radiation) which randomly flip bits in semiconductor devices.

%Furthermore, vulnerabilities in modern DRAM density can be exploited by security threats such as Rowhammer attacks, where frequent row activation induces bit-flips in adjacent memory rows, effectively simulating SDC behavior through intentional interference.

\subsection{LLMs Training at Scale}
\label{sec:sec:llm}
%\fixme{Give readers an idea of the scale of llm training and how hardware fault rate has gone up with this scale. Cite recent meta work~\cite{kokolis2025revisiting}}.
The training of Large Language Models (LLMs) has reached an unprecedented scale, necessitating distributed systems that span thousands of accelerators to handle models with hundreds of billions of parameters. State-of-the-art models require massive computational resources over extended periods; for instance, training Llama 3 405B utilized a cluster of 16,000 H100 GPUs~\cite{metallama3_2024}, and other large-scale efforts similarly employ tens of thousands of chips. This massive parallelism typically combines data, tensor, and pipeline parallelism to partition the workload, so that a fault in a single node can perturb the global training run.
As the number of hardware components in a training cluster increases, the aggregate probability of encountering hardware failures, including SDC, rises significantly~\cite{Jorg:2025}. Empirical reports from major industry training runs indicate that SDC events are no longer rare anomalies but regular occurrences at scale: Meta attributed six unplanned job interruptions to SDC during a 54-day pre-training snapshot~\cite{metallama3_2024}, and Google estimated that an SDC event occurs every week or two during Gemini training~\cite{gemini2023gemini}. Compounding this issue, modern high-performance LLM training workloads push hardware utilization to its limits, which can activate latent defects in marginal hardware that would likely pass manufacturing screening stress tests~\cite{cui2025characterizing}.

\subsection{Intuitions About LLM Fault Resilience}
\label{sec:sec:intuition}
A widely held intuition is that LLM training may be inherently resilient to transient hardware errors: the optimization process is stochastic by design, gradient clipping and norm-based regularization act as implicit error dampeners, and the sheer volume of gradient updates across billions of parameters can dilute the impact of isolated corruptions. This intuition has merit for occasional transient faults, but it does not necessarily extend to permanent and intermittent hardware faults, where errors recur systematically across training steps rather than appearing as isolated perturbations. Given the scale and economic significance of large-scale LLM training infrastructure, characterizing the degree to which such permanent faults degrade convergence and destabilize training dynamics is therefore necessary to reason rigorously about the resilience properties required of LLM training.

% \fixme{give the citations here in the table. also i thought your focus was pre-training as well from the abstract? what's the point of the last column? it seems that you are the worst there. also do you want to emphasize that you have two architectures and three data types but prior work didn't?}
\begin{table*}[ht]
\centering
\caption{Comparison of prior LLM fault-tolerance and resilience studies. We contrast studies along key dimensions including hardware realism, fault type, training phase, injection scope, distributed support, and scale. For our work, total compute time is estimated from 7,664 complete training runs by weighting model training time by the number of runs per model. Sun et al.~\cite{sun2025demystifying} and Yu et al.~\cite{yu2025exploring} report total hours explicitly; Ma et al.~\cite{ma2025understanding} and Chen et al.~\cite{chen2025bit} do not report total GPU hours, so their reported experimental bounds are listed instead.}
\label{tab:related_work_comparison}
\resizebox{\textwidth}{!}{%
\begin{tabular}{@{}lcccccccc@{}}
\toprule
\textbf{Work} & \textbf{\begin{tabular}[c]{@{}c@{}}Hardware\\ Accurate\end{tabular}} & \textbf{Fault Type} & \textbf{\begin{tabular}[c]{@{}c@{}}Ease of\\ Exp.\end{tabular}} & \textbf{\begin{tabular}[c]{@{}c@{}}Target\\ Phase\end{tabular}} & \textbf{\begin{tabular}[c]{@{}c@{}}Injection\\ Scope\end{tabular}} & \textbf{\begin{tabular}[c]{@{}c@{}}Analyzes\\ Dynamics\end{tabular}} & \textbf{\begin{tabular}[c]{@{}c@{}}Distributed\\ Support\end{tabular}} & \textbf{\begin{tabular}[c]{@{}c@{}}Total Compute \\ / Scale\end{tabular}} \\ \midrule

Ma et al.~\cite{ma2025understanding} & \checkmark & Real / Latent & $\times$ & \begin{tabular}[c]{@{}c@{}}Pre-train \& \\ Fine-tune\end{tabular} & Submodules, Gradients & \checkmark & \checkmark & \begin{tabular}[c]{@{}c@{}}Not reported\\ (30 nodes, 4.5k steps)\end{tabular} \\ 

Chen et al.~\cite{chen2025bit} & $\times$ & Transient & \checkmark & \begin{tabular}[c]{@{}c@{}}Fine-tune \& \\ Inference\end{tabular} & Weights, Activations & $\times$ & $\times$ & \begin{tabular}[c]{@{}c@{}}Not reported\\ (5 epochs/run)\end{tabular} \\ 

Sun et al.~\cite{sun2025demystifying} & $\times$ & Transient & \checkmark & Inference Only & Weights, Neurons & $\times$ & $\times$ & \begin{tabular}[c]{@{}c@{}}$\sim$4,800 hours \\ (13M+ runs)\end{tabular} \\ 

Yu et al.~\cite{yu2025exploring} & $\times$ & Transient & \checkmark & Training & \begin{tabular}[c]{@{}c@{}}Forward, Gradients, \\ Optimizer states\end{tabular} & \checkmark & \checkmark & \begin{tabular}[c]{@{}c@{}}$>$5,000 hours \\ (300K+ runs)\end{tabular} \\ \midrule

\textbf{Our Work} & \textbf{\checkmark} & \textbf{\begin{tabular}[c]{@{}c@{}}Permanent\end{tabular}} & \textbf{\checkmark} & \textbf{Training} & \textbf{\makecell{Weights, Inputs,\\ Gradients}} & \textbf{\checkmark} & \textbf{\checkmark} & \textbf{\begin{tabular}[c]{@{}c@{}}$\approx$11,500 hours \\ (7,664 runs)\end{tabular}} \\ \bottomrule
\end{tabular}%
}
\end{table*}

\subsection{Related Work}
\label{sec:sec:related}
%\fixme{Describe previous work when it comes to studying hardware fault and its impact on the overall networks. This involves works on CNNs used in AVs etc. Then mention that with use of LLMs, there has been recent work that looks at LLM inference under faults. But there aren't a lot of works that have looked at impact of faults during LLM training. Mention the ACL works~\cite{ma2025understanding,chen2025bit} and SC work~\cite{yu2025exploring,sun2025demystifying}. Our work is different as we provide you with a way to do characterize models without running them on actual faulty nodes}.

Table~\ref{tab:related_work_comparison} exposes the key limitations of prior LLM fault-tolerance and resilience studies, particularly their reliance on idealized fault models, short evaluation horizons, and limited coverage of end-to-end training dynamics.
These studies differ in fault realism, evaluation horizon, and whether they capture end-to-end training dynamics. Idealized bit-flip injections, short-horizon evaluations, and inference-only studies can miss failure modes that emerge with a realistic hardware-grounded fault model and long-running pre-training.

Sun et al.~\cite{sun2025demystifying} present a large fault-injection study, but focus only on inference. Yu et al.~\cite{yu2025exploring}, Chen et al.~\cite{chen2025bit}, and Ma et al.~\cite{ma2025understanding} extend the scope to training, but remain limited by either short fine-tuning horizons or reliance on physically degraded hardware, which constrains reproducibility and the range of fault signatures that can be explored, and the range of training trajectories that can be observed. In particular, prior training studies are often too short to expose long-tail phenomena such as delayed loss spikes, divergence, or silent corruption that only become visible over extended runs.

Our work addresses these gaps by combining hardware-grounded fault models with software-based injection inside the training loop, enabling large-scale pre-training studies from initialization to convergence. This allows us to study not only average degradation, but also loss spikes, NaN/Inf anomalies, and divergence across the full optimization trajectory of LLM training.

\section{Methodology}
\label{sec:method}

% This section presents our methodology for characterizing the impact of permanent and intermittent hardware faults on LLM training. The challenge we face is that of fidelity vs scale: cycle-accurate RTL simulation of hardware faults accurately captures defect behavior but is computationally prohibitive for end-to-end training runs~\cite{}, while naive software-level bit-flip injection is fast but risks producing error patterns that do not correspond to any realistic hardware failure mode~\cite{}. 

The methodology first situates our approach in the fault-injection design space by contrasting RTL-level and model-level injection (Section~\ref{sec:method:model}). It then derives hardware-grounded error signatures from representative GPU functional units under realistic stuck-at and timing-dependent defects (Section~\ref{sec:method:rtl}). These signatures drive a software-level fault-injection framework embedded in the training pipeline and parameterized by a seven-element fault-site tuple that specifies where, when, and how faults occur (Section~\ref{sec:method:fault}). Finally, a Bernoulli error-rate parameter models intermittent activation and unifies permanent and intermittent faults within a single probabilistic framework (Section~\ref{sec:method:error_rate}).

%\subsection{Permanent and Intermittent Fault Model}
%\label{sec:method:faultmodel}
% \paragraph{Intermittent Faults:}This work develops a fault model that captures the behavioral signatures of permanent and intermittent hardware faults rooted in well-understood physical degradation mechanisms. Permanent faults can occur due to manufacturing test escapes when defective devices pass production screening and only reveal themselves under field conditions, latent early-life defects that surface during the initial operational period as marginal structures fail under stress, and due to aging-induced failures develop progressively over a chip's lifetime as physical wear-out mechanisms such as BTI, HCI, TDDB, and EM gradually degrade device and interconnect characteristics.

% At the logical level, permanent faults may manifest as stuck-at faults in combinational logic, flip-flops, or RAM arrays, where a node is permanently fixed at a logic '0' or '1' regardless of the applied input — these are timing-independent faults whose effect is deterministic and persistent across all operating conditions. Partial degradation of transistors or interconnects introduces small delay defects that reduce the available timing margin in a circuit path. When these delays grow sufficiently large, they cause violations of setup or hold time constraints, leading to incorrect values being latched or propagated. Such faults are classified as timing-dependent permanent faults, or intermittent faults, exhibit marginality - they tend to manifest selectively at specific voltage or frequency operating points.

\subsection{From Physical Defects to Logical Faults}
\label{sec:method:model}

Permanent GPU defects, including intermittently activated ones under varying thermal and electrical stress, are a key source of SDC during LLM training. To study them systematically, we map physically plausible defect mechanisms to logical perturbations in the training pipeline. Rather than simulate faults at the hardware level, we simulate their observable error symptoms using statistical activation models that capture when and how errors appear over time. This approach preserves the realism of permanent and intermittent defects while enabling scalable end-to-end evaluation. We instantiate it in Megatron-LM, a production-grade training framework widely used for transformer models on GPUs.

\paragraph{Assumption:} In this work, a single underlying hardware defect is modeled at a time, i.e., at most one permanent fault is assumed to be active in the system over the duration of each training run; this simplifies attribution of observed behavior to a specific defect and reflects the fact that, at realistic fault rates, multi-defect interactions are relatively rare and can be studied in future work.

% \begin{figure*}[ht]
%     \centering
%     \includegraphics[width=\textwidth]{figs/fig_method.pdf}
%     \caption{Intermittent Fault Characterization Methodology}
%     \label{fig:method}
% \end{figure*}

\subsection{RTL Fault Characterization}
\label{sec:method:rtl}

\begin{figure}[t]
    \centering
    \includegraphics[width=\columnwidth]{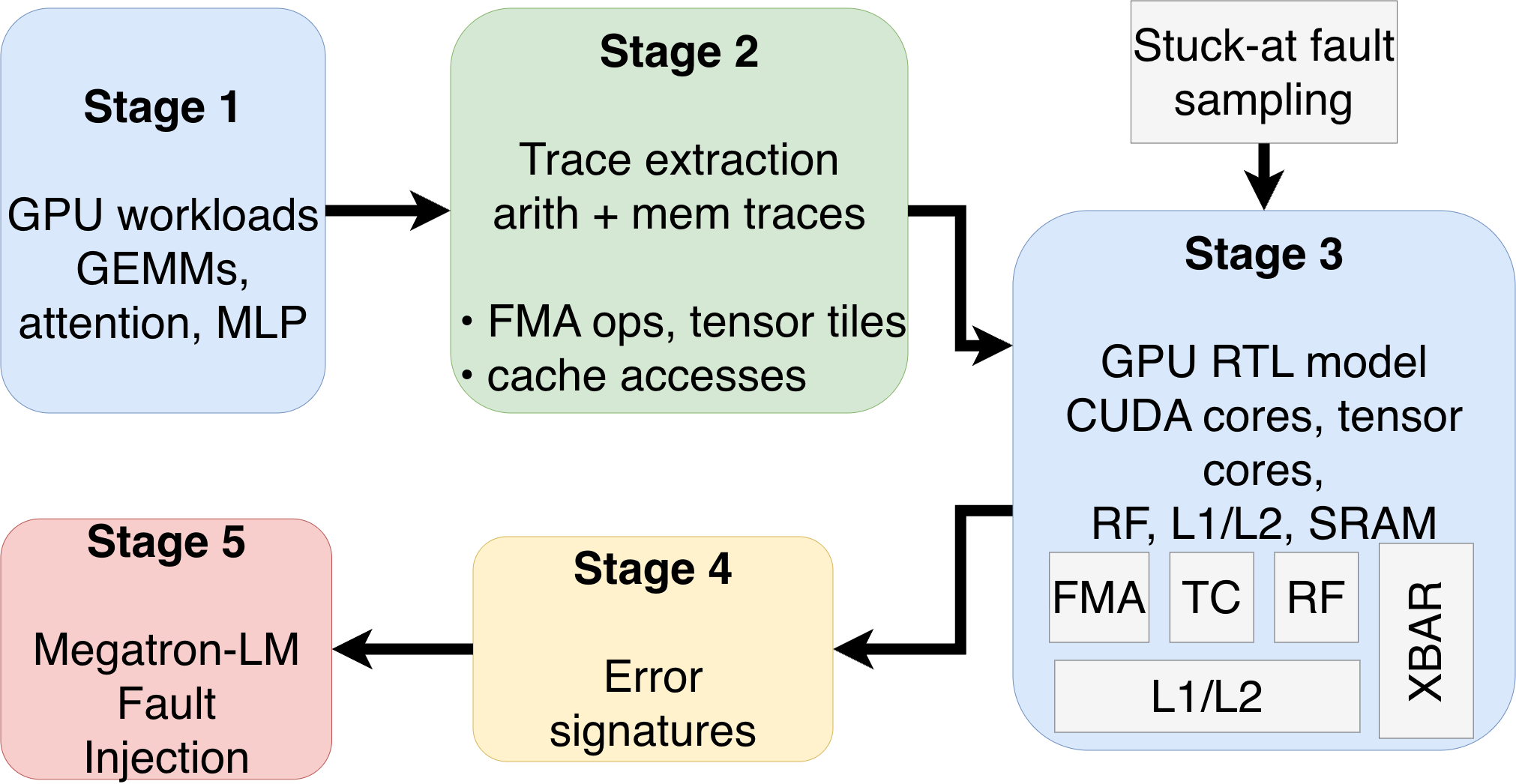}
    \caption{RTL characterization flow for error signature extraction.}
    \label{fig:error_signatures}
\end{figure}

\begin{figure*}[ht!]
    \centering
    \includegraphics[width=0.95\textwidth]{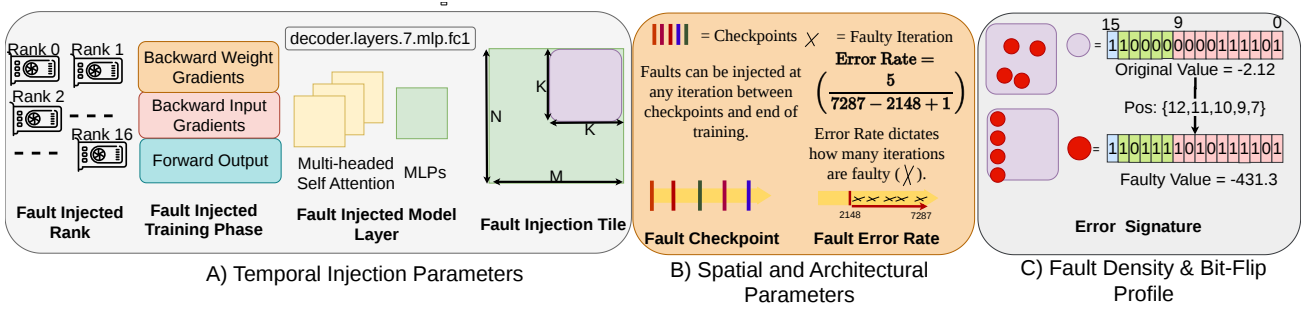}
    \caption{Software level fault site tuple characterized with seven parameters divided in three broad categories: 1) Temporal 2) Spatial and Architectural 3) Fault Intensity \& Bit-Level.}
    \label{fig:fault_site}
\end{figure*}
Figure~\ref{fig:error_signatures} summarizes our RTL characterization flow. Rather than simulate the full LLM training workload with an RTL simulator --- which is computationally intractable~\cite{Hukerikar:DSN:2024} --- we target the GPU functional units most exposed during transformer training: FMA paths in the CUDA cores, tensor cores, register files, and L1/L2 caches and SRAM arrays. Simulations are driven by computation traces extracted directly from the training workload and replayed through a commercial EDA tool to obtain cycle-accurate functional-unit activity under realistic input and toggle conditions.

We inject stuck-at faults sampled in proportion to the silicon area of each unit, ensuring the fault population reflects each block's contribution to the overall GPU design. For each fault, we record its output error signature: the set of affected bits, the magnitude of the numerical deviation, and the corruption pattern at the functional-unit output. These signatures capture the observable software-level manifestation of a fault after propagation through the GPU's actual modeled datapaths, control logic, scheduling behavior, and RAS mechanisms. The signatures characterize the \textit{what} of each fault --- which bits are affected, at what magnitude, and in what pattern --- while the \textit{where} and \textit{when} are determined by architectural reasoning about GPU data flow and training dynamics, as described in next subsection. 

\begin{figure}[ht!]
    \centering
    \includegraphics[width=0.95\columnwidth]{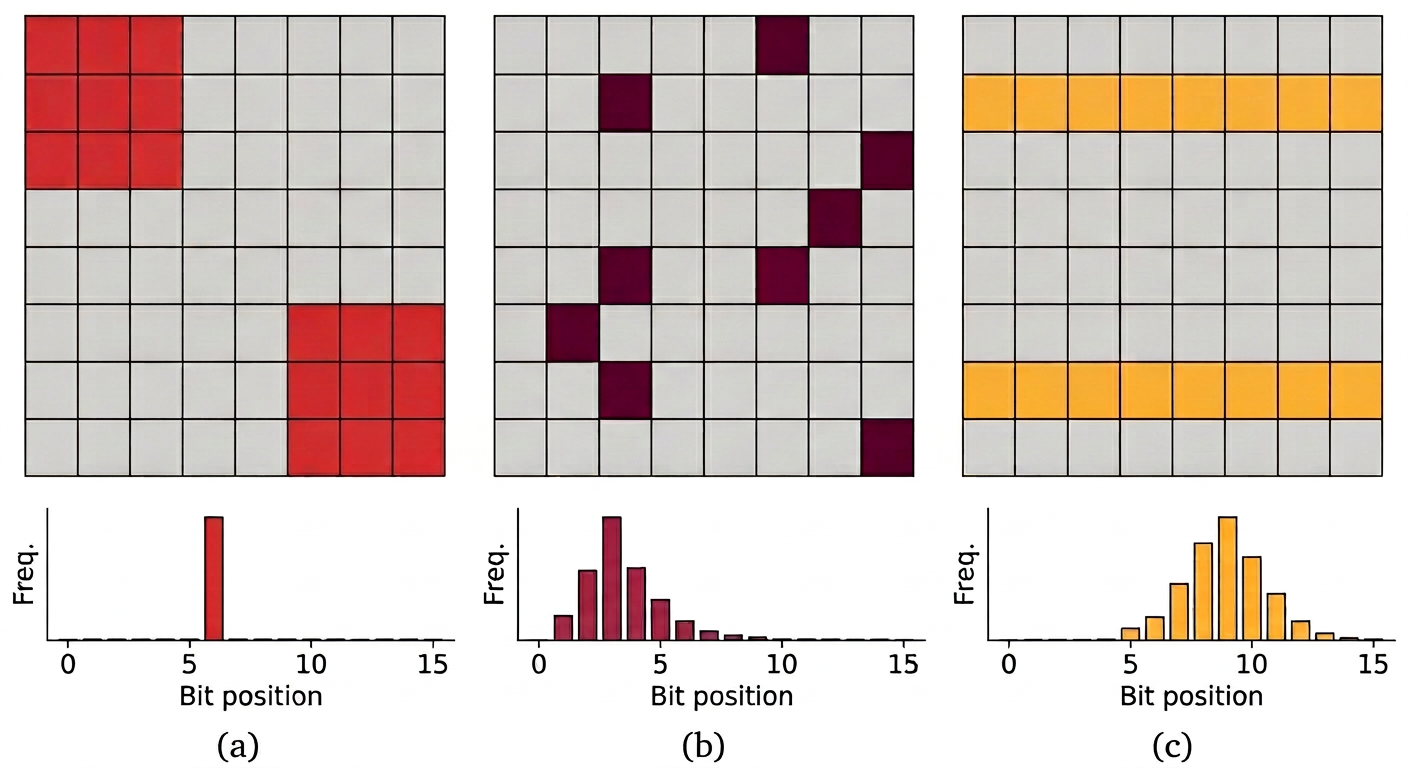}
    \caption{Example error signatures derived from RTL characterization.}
    \label{fig:rtl_sign}
\end{figure}

The error signatures in Figure~\ref{fig:rtl_sign} are illustrative examples derived from single stuck-at fault RTL simulations. For Figure~\ref{fig:rtl_sign}(a), the fault is injected into a tensor core datapath bit that feeds a subset of output tiles, producing localized $3\times3$ patches in the matrix where all corruptions share the same flipped bit position, resulting in a sharply peaked bit-position distribution. For Figure~\ref{fig:rtl_sign}(b), the fault is injected on an output bit of an FMA ALU, perturbing low- and mid-order bits in a pattern-dependent manner so that most outputs remain correct but a small number of operations exhibit sporadic, high-magnitude numeric outliers with a broader bit distribution. In case (c), the stuck-at fault is in L1 cache line such that an entire matrix row’s stored values are corrupted when accessed, yielding structured row-wise corruption and a wide bit-position distribution centered on high-order bits. 

The RTL fault-injection platform is derived from the implementation RTL corresponding to the same NVIDIA H100 GPU, which is used for the large-scale software-based fault injection experiments. Consequently, the microarchitectural blocks targeted by injection — including the relevant datapaths, control logic, storage structures, and reliability, availability, and serviceability (RAS) mechanisms — represent the same logical hardware structures instantiated in silicon for the H100 GPU used for the software-based fault injection. This correspondence allows the software-level fault-injection experiments to be interpreted against the behavior of the deployed GPU under the same architectural and microarchitectural configuration. In particular, the injected RTL faults exercise the design's actual error-propagation paths and protection mechanisms rather than those of an abstracted or functionally simplified GPU model. We collected over 6,000 such error signatures from faults injected throughout the GPU memory hierarchy and compute datapaths providing a diverse empirical basis for the software-level fault injection.

%\section{Software Level Intermittent Fault Injection Framework}
%\label{sec:soft}

%\subsection{Software Fault Site Tuple}
\subsection{Software Fault Site Parameterization}

RTL fault injection operates on the registers and combinational data paths of a functional unit; the analogous injection targets in a software training loop are the tensors flowing through the computation graph --- activations, gradients, and weights. Translating hardware-derived error signatures into this setting requires a structured parameterization that captures the
spatial, temporal, and numerical dimensions of a physical defect as quantities controllable within Megatron-LM.

The core challenge is that a single stuck-at defect has a fixed physical location, but its effect on training depends on which tensor is being computed, on which device, at which step, and at what numerical precision. A parameterization that spans all of these dimensions lets us reproduce the full diversity of hardware failure modes during distributed training with Megatron-LM.

To this end, we define a software-level fault site as a tuple of seven parameters organized into three categories (Figure~\ref{fig:fault_site}). The fault intensity and bit-level parameters are drawn from the RTL-derived error signature distributions established above; the spatial and architectural
parameters follow from first-principles reasoning about how a fixed hardware defect interacts with distributed training computation; and the temporal parameters capture the stochastic activation behavior of intermittent defects, as detailed in Section~\ref{sec:method:error_rate}. The seven parameters are described below: 

\label{sec:method:fault}

\subsubsection{Spatial and Architectural Coordinates}
\label{sec:sec:spatial}
These three parameters specify which hardware and which part of the model is affected (Figure~\ref{fig:fault_site}b).

\textbf{a) Rank} identifies the faulty GPU in the distributed cluster, simulating the common datacenter scenario where a single unhealthy device operates alongside otherwise healthy peers.

\textbf{b) Layer} pins the fault to a specific transformer layer and sub-module --- Multi-Head Attention, MLP, or LayerNorm --- enabling study of whether certain components are more susceptible to corruption than others.

\textbf{c) Training Phase} determines which tensor is corrupted: forward-pass activations (\texttt{fwd\_outputs}), backward-pass input gradients (\texttt{bwd\_grad\_inputs}), or weight gradients (\texttt{bwd\_grad\_weights}). The distinction is consequential: an activation error is consumed and discarded, whereas a weight gradient error propagates directly into the model parameters,
causing persistent drift in the optimization trajectory.

\subsubsection{Temporal Injection Parameters}
\label{sec:sec:temporal}
These two parameters specify when and how often corruption occurs during training.

\textbf{a) Fault Checkpoint} sets the iteration at which injection begins, enabling controlled comparisons of model vulnerability at different stages of convergence --- early training, when gradients are large and noisy, versus later stages, when the model is near a local minimum and more sensitive to perturbation.

\textbf{b) Error Rate} controls the fraction of eligible iterations that experience a fault, spanning the full persistence spectrum: a rate of 1.0 models a permanent stuck-at defect corrupting every iteration, while a low rate models an intermittent defect that activates under specific input or operating conditions. Section~\ref{sec:method:error_rate} describes how this parameter is derived and swept.

\subsubsection{Fault Intensity \& Bit-Level Mechanics}
\label{sec:bitlevel}
These two parameters specify the severity and structure of the numerical corruption applied to the target tensor.

\textbf{a) Fault Density} controls the proportion of corrupted elements within the target tensor tile. A low density models a localized defect; a higher density models a fault in shared logic --- such as a cache line or register file entry --- that corrupts multiple values simultaneously.

\textbf{b) Bit-Flip Profile} specifies which bits of the floating-point representation are flipped and how many. Flipping an exponent bit can shift a value by orders of magnitude (e.g., $3.141 \rightarrow 3248$), whereas mantissa flips produce minor precision loss. By replaying the bit-flip patterns observed in the RTL error signatures, this parameter anchors the software fault model in hardware-measured corruption rather than uniform random bit-flip assumptions.

\subsection{Instantiating a Fault Site: Fixed and Variable Parameters}
\label{sec:sec:choose_fault}

%\fixme{This section justifies why faults are chosen in this manner and what is the hardware behavior of the corresponding fault and their decisions.}

% We describe how the fault injection parameters shown in \Fig{fig:method} are chosen for a fault injection campaign. This is unique to the intermittent faults we are trying to emulate because in a single campaign, the same fault can have different behavior depending on the input data. Hence, a fault chosen at a single iteraiton may or may not be the same as a different one. This means that out of all the parameters of our fault tuple, we have to be precise about what values in that tuple remains the same or changes across different faulty iterations (marked with $\times$ in \Fig{fig:method}) and make sure it models the hardware fault behavior correctly. 
%Not all parameters of the fault site tuple change between faulty iterations within a single injection campaign. A physical hardware defect is fixed in one location — but depending on what data flows through the defective circuit, it may or may not produce a visible error, and the error it produces may affect different layers or training phases. This distinction between what is physically fixed and what varies with the data stream determines which tuple parameters remain constant and which are resampled at each faulty iteration during software level fault injection. 

Instantiating the fault-site tuple for a concrete injection campaign requires deciding which of its seven parameters reflect fixed properties of the physical defect and which must be resampled as the data stream changes across iterations. A physical defect is anchored to one location on one device, but the tensor it corrupts, the layer it affects, and the bit pattern it produces all depend on what computation happens to be scheduled on the defective unit. This distinction between defect-level properties and data-stream interactions determines which parameters are initialized once at
campaign start and which vary throughout training.

\subsubsection{Fixed Parameters}
\label{sec:sec:const}

\textbf{a) Rank} is fixed because a hardware defect resides on a specific physical device. A degraded functional unit on GPU~$k$ remains faulty for the entire training run, and we assume at most one faulty GPU at any time.

\textbf{b) Fault Checkpoint} and \textbf{c) Error Rate} are fixed because they describe the temporal characteristics of the defect rather than any individual corruption event. The checkpoint defines the training window during which the fault is active --- modeling, for example, a defect that emerges after thermal stress accumulates over extended operation. The error rate captures how frequently the defect propagates to a visible output error within that window. 

\subsubsection{Variable Parameters}
\label{sec:sec:var}

\textbf{a) Training Phase} is sampled uniformly from
\texttt{fwd\_outputs}, \texttt{bwd\_grad\_inputs}, and
\texttt{bwd\_grad\_weights}. This is an architectural choice: the forward pass performs one matrix computation per layer, whereas the backward pass performs two (input and weight gradients), so a uniform draw over three phases yields a natural 2:1 exposure ratio for backward versus forward computation without explicit timing measurements.

\textbf{b) Layer} is sampled uniformly across transformer layers. Because layers are structurally identical and occupy the same GPU for comparable durations, uniform sampling reflects equal exposure of a fixed defect to each layer's computation.

\textbf{c) Fault Injection Tile} is sampled to match GPU execution granularity. Matrix operations are executed in fixed-size tiles; a defect corrupts whatever tile is scheduled on the faulty unit at activation. We therefore select a target tile at random and derive the intra-tile corruption pattern from the RTL error signature distributions.

\textbf{d) Bit-Flip Profile} is drawn from the RTL error signatures with single-bit flips weighted more heavily than multi-bit patterns. 

% \begin{algorithm}[ht]
% \SetAlgoLined
% \KwIn{Model $M$, Total Steps $T$, Checkpoint Window $fc$, Error Rate $er$, Rank $r$, Mode $mode \in \{\text{RTL}, \text{MC}\}$}
% \KwOut{Trained model $M'$ with fault injection}

% \tcp{Phase 1: Initialize Constant Parameters (Before Training)}
% Initialize fault campaign with static parameters: $(fc, er, r)$\;
% \If{$mode == \text{RTL}$}{
%     Load RTL Error Signatures Database $\mathcal{D}_{RTL}$\;
% }

% \tcp{Phase 2: Active Training Loop}
% \For{step $t \leftarrow 1$ \KwTo $T$}{
%     \If{$t \in$ window defined by $fc$}{
%         Draw $u \sim \text{Uniform}(0,1)$ \tcp*{Check Intermittent Activation}
%         \If{$u < er$}{
%             \tcp{Sample Variable Parameters (Spatial \& Architectural)}
%             Sample Phase $p \sim \text{Uniform}(\{\text{fwd}, \text{bwd\_inp}, \text{bwd\_wt}\})$\;
%             Sample Layer $l \sim \text{Uniform}(1, L_{total})$\;
%             Sample Destination Tile $k$\;

%             \tcp{Determine Fault Intensity \& Mechanics}
%             \eIf{$mode == \text{RTL}$}{
%                 Draw signature $(f, b) \sim \mathcal{D}_{RTL}$ based on tile $k$ and phase $p$\;
%             }{
%                 Draw \% of faults $f$ and bit-flips $b$ via Monte Carlo sampling\;
%             }

%             \tcp{Apply Software-Level Fault}
%             Inject fault at $(r, l, p, k, f, b)$ into tensor\;
%         }
%     }
%     Execute training step $t$ on model $M$\;
% }
% \Return $M'$\;
% \caption{Two-Stage Intermittent Fault Injection Campaign}
% \label{alg:fault_injection}
% \end{algorithm}

\subsection{Modeling Intermittent Faults With Error Rate}
\label{sec:method:error_rate}
%\fixme{This section will describe the problem with doing software level fault injection. Then it will introduce error rate as a solution to it, describing properly the point of it and its use case.}
\begin{figure}[t]
    \centering
    \includegraphics[width=0.7\columnwidth]{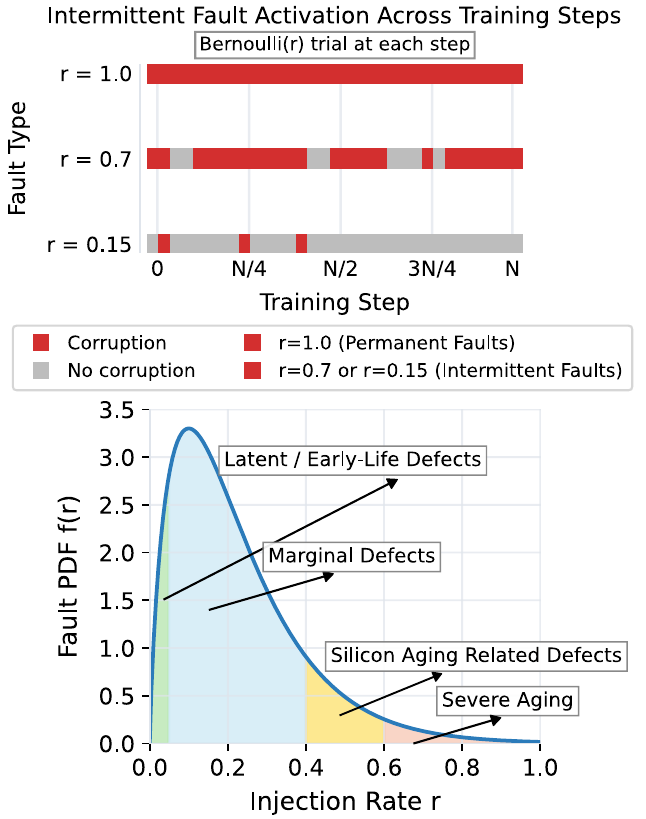}
    \caption{Stochastic Intermittent Fault Activation Model}
    \label{fig:intermittent_fm}
\end{figure}

To model intermittent faults — permanent hardware defects that only manifest on a subset of operations — we interpret the error-rate field in the fault tuple as the probability that a given spatial fault site produces a visible corruption during training. Whether a given operation produces a visible error depends on whether the input data exercises the defective logic path, or whether the operating conditions — voltage, frequency, or thermal stress — cross the threshold at which the defect becomes active. This combination of data-dependent and condition-dependent activation is what makes permanent faults behave intermittently at the software level.

The \emph{error injection rate} \(r \in (0,1]\) is defined for a given defect as the per-iteration probability that the defect produces a visible corruption at the software level, conditioned on the operation using the faulty resource. In other words, \(r\) summarizes the defect’s \emph{detectability}: the fraction of input patterns that exercise the faulty logic path and propagate an error to architecturally visible state. Each fault tuple is associated with a fixed rate \(r\) for the duration of an injection campaign. At each eligible iteration, a Bernoulli trial with parameter \(r\) decides whether the defect activates; on activation, the fault tuple is used to inject the error to the target tensor. A rate \(r = 1.0\) models a hard stuck-at defect that corrupts every eligible operation, whereas \(r \ll 1\) models a timing‑marginal or latent defect that rarely manifests. 

To study how activation frequency shapes training behavior, we treat \(r\) as a hyperparameter and sweep it across a broad range. Figure~\ref{fig:intermittent_fm} visualizes the fault activation over training steps as a Bernoulli process with rate \(r\), and the the corresponding probability density function \(f(r)\). We construct PDF $f(r)$ based on prior reliability reports: these studies consistently observe that most permanent or intermittent defects activate rarely, while a smaller fraction exhibit much higher activation rates. We therefore choose a unimodal distribution with most probability mass concentrated at low \(r\)
(latent and marginal defects), but with a long tail covering higher 
\(r\) values corresponding to aging and severely degraded structures. For our experiments, we select a set of representative rates spanning these regions and run separate campaigns at each value, so that each campaign probes a distinct point in this error activation rate design space. This ties the error rate parameter back to the physical intuition about intermittent defects while providing a systematic knob for characterizing how faults of different activation frequencies affect LLM training stability~\cite{Saxena:EM:2022}.

% \textbf{RQ4: What segments of the training are more susceptible to hardware faults?} \\

% \textbf{RQ5: How are NaN/Inf dealt with during the model training?} \\

% \textbf{RQ6: What is the most dominant factor in the fault injection tuple that determines the impact of hardware faults on models?} \\

% \textbf{RQ7: What is the impact of fault happening multiple times during training?}

% \textbf{RQ: How does a fault travel from the start to the end of training?}
\section{Experiment Setup}
\label{sec:exp}

% \subsection{Setup}
% \label{sec:exp:setup}
%\fixme{Describe the details of our experimental section. This includes talking about the models, dataset and the number formats we chose. We also need to describe why we don't do more experiments on different models + datasets. Also describe the settings in Megatron that we do along with the hardware we chose to do fault injection emulations.}

\subsection{Models and Dataset}
\label{sec:exp:model}

%\paragraph{Models:}We choose the GPT-2~\cite{radford2019language} architectures to do the fault injection experiments. To see the impact of model scaling, we carry out experiments on both the small (122M) and Medium (345M) variants of the model. 
%All the experiments are carried out with WikiText~\cite{} dataset and we measure the PPL of all the runs to see the impact of fault injections at the macro level. 

%We choose a decoder based architecture as GPT-2 serves as the foundational archetype for the decoder-only transformer architectures that dominate the current LLM landscape (e.g., Llama~\cite{metallama3_2024}, GPT-4~\cite{achiam2023gpt} etc.).

%Moroever, to achieve statistical significance when evaluating intermittent faults over complex optimization landscapes, a high volume of independent training runs is required. We chose to limit our evaluation to two GPT-2 architectures (small and medium) trained on the Wikitext dataset to prioritize the depth and completeness of the training trajectory over model scale. Specifically, we conduct 7,664 (2981 for GPT-2Medium and 4681 for GPT-2Small) complete training experiments from scratch to convergence.

%Evaluating larger, contemporary models for thousands of full pre-training lifecycles is computationally prohibitive. For context, we compare the scale of our studies to other works in the literature and provide a summary in ~\Tbl{tab:related_work_comparison}.

\paragraph{Models:} We perform all fault-injection experiments on the GPT-2 family of decoder-only Transformers~\cite{radford2019language}, using the 124M (GPT-2 Small) and 355M (GPT-2 Medium) variants. These models represent the foundational archetype for modern decoder-only LLMs (e.g., Llama, GPT-4), while remaining small enough to support thousands of full training runs. To study the effect of model
scaling under realistic training dynamics, we train both variants from scratch on the WikiText language modeling dataset~\cite{merity2016pointer}. This experimental design deliberately trades absolute model size for statistical rigor: by restricting to two GPT-2 architectures, we are able to run 7,664 complete fault-injected pre-training experiments, enabling fine-grained evaluation of a large space of
fault sites and activation rates.

\paragraph{Dataset:}All the experiments are carried out with WikiText~\cite{merity2016pointer} dataset and we measure the PPL of all the runs to assess the impact of fault injections at the macro level. 
A rigorous fault injection campaign requires thousands of targeted training iterations to achieve statistical significance across a massive configuration space (layer depth, error rate, data format etc). By utilizing a single, sufficiently complex dataset that thoroughly exercises the model's computational graph, we eliminate dataset-induced variance as a confounding factor while keeping the computational overhead of exhaustive fault profiling tractable.

\subsection{Data Formats}
\label{sec:exp:data}
To evaluate the impact of hardware faults across modern mixed-precision regimes, we run all fault-injection campaigns under three training formats: IEEE FP16, BF16~\cite{kalamkar2019study}, and FP8~\cite{micikevicius2022fp8}, which are the de facto low-precision standards. Rather than varying datasets or model sizes, we fix the architecture and data and sweep the numerical format, since the low-level propagation of a bit-flip through PyTorch primitives is governed primarily by the underlying representation. For example, flipping an exponent bit in an FP16 tensor can change a value by orders of magnitude, whereas the same spatial fault in a BF16 tensor, with a wider exponent and narrower mantissa, produces a different corruption profile; FP8 further alters this trade-off by aggressively compressing both range and precision. While increasing model scale is known to affect activation outlier distributions and thus quantization behavior~\cite{dettmers2022gpt3}, isolating the numerical format at a controlled scale allows us to rigorously probe a large space of hardware-level bit-flips and activation rates, yielding more direct, actionable insight into how representation choice shapes training-time fault sensitivity.

\subsection{Hardware, Software and Training Recipe}
\label{sec:exp:hwswtrain}

%\fixme{Make sure the details are correct.}

\paragraph{Hardware:} The fault injection training experiments were run on 16× NVIDIA H100 GPUs spread across two 8-GPU nodes, using data-parallel Megatron-LM training. Each experiment occupies all 16 GPUs across both nodes.
    
\paragraph{Software:} We utilize the open-source Megatron-LM~\cite{shoeybi2019megatron} framework, which we specifically modified to integrate our targeted fault injection methodology during model training.
    
\paragraph{Training Recipe:} We employ a standard configuration using the Adam optimizer, linear warmup with cosine annealing, and sequence parallelism for distributed workloads. 
%Comprehensive details regarding the training hyperparameters and scaling configurations are provided in the appendix. 
Across all configurations, this setup yields 7{,}664 complete fault-injected pre-training runs from scratch (4{,}681 for GPT-2 Small and 2{,}983 for GPT-2 Medium), in which RTL-derived error signatures are used to construct the fault-site tuples and the error-rate parameter is swept across its range.

\paragraph{RTL-Derived Software Fault-Injection Model:} The software-level injection campaign uses only signatures derived from faults that evade on-chip RAS detection and reach architecturally visible tensor values as silent data corruption. We exclude faults that are masked, corrected, or escalated into a detected error, because those faults do not represent silent corruption of the training computation under the modeled execution conditions. Thus, our software injection does not use arbitrary independent bit flips, random noise, or synthetic tensor corruptions. Instead, it replays the locality, shape, affected-element set, and persistence properties measured in RTL for faults that produce undetected tensor-value corruption.
This distinction is important because a hardware fault need not manifest as a single independent bit flip at the software interface. The observed corruption pattern can be structured and spatially correlated across tensor elements for a permanent fault in the hardware.

\section{Results}
\label{sec:results}
%\label{sec:res:rtl}

\begin{figure*}[ht!]
    \centering
    \includegraphics[width=\linewidth]{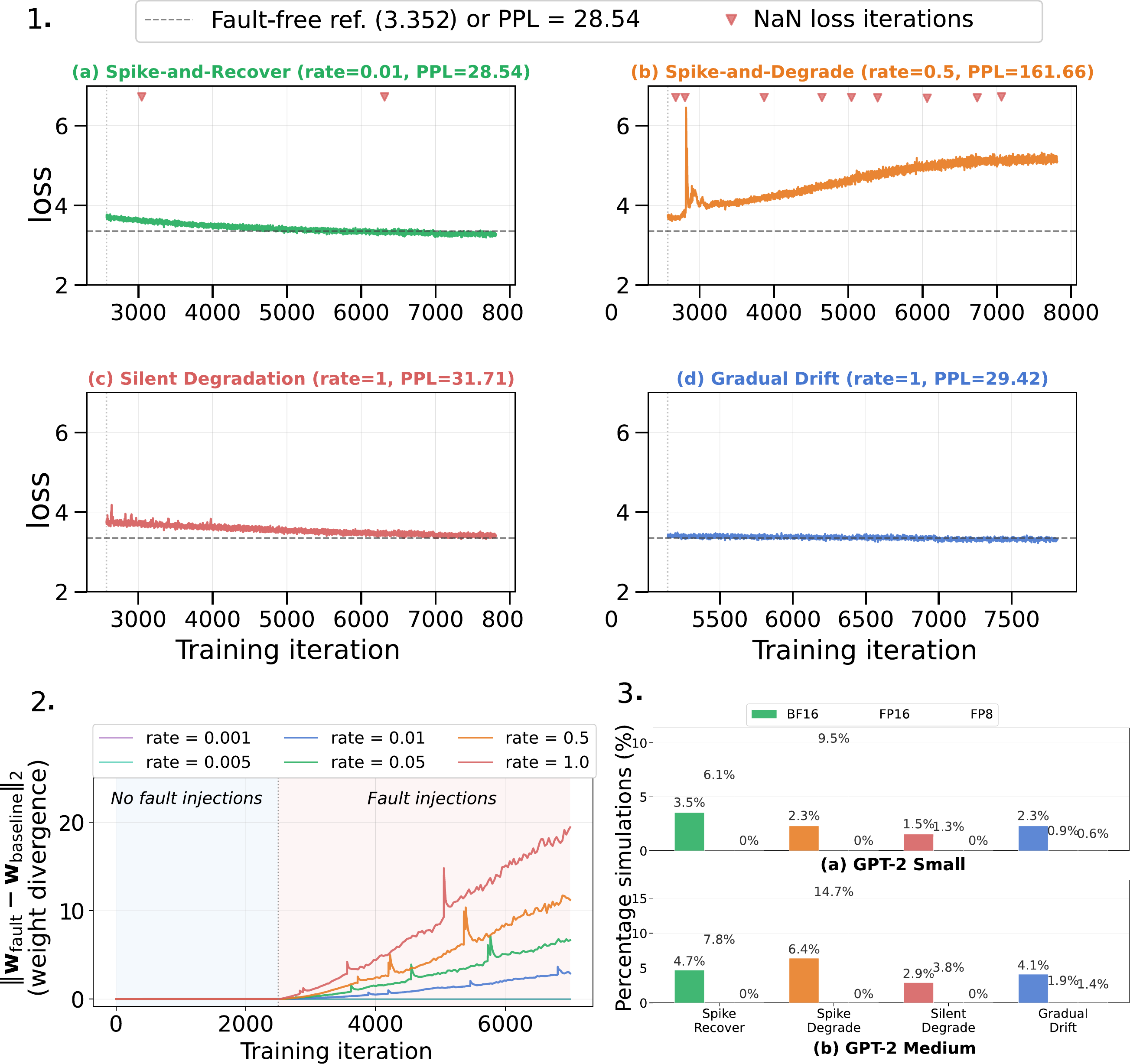}
    \caption{Training loss traces for four representative failure modes under permanent fault injection. Each subplot shows loss over training iterations (GPT2-Small with BF16 format); the dashed line is the fault-free baseline (3.352 nats / PPL = 28.54) and red triangles mark iterations where the loss calculation produced a NaN. \textbf{Plot 1:}(a)Spike-and-Recover: sparse NaN events are absorbed and the run converges to baseline. (b)Spike-and-Degrade: dense NaN events overwhelm recovery and loss diverges. (c)Silent Degradation: no NaN events, yet loss settles above baseline; the most operationally deceptive mode. (d)Gradual Drift: loss is visually indistinguishable from baseline while parameters are silently corrupted. \textbf{Plot~2:} Global weight divergence ($\|W_{\text{fault}} - W_{\text{baseline}}\|_2$) over training, for GPT-2Small with BF16. All fault rates produce monotonically increasing divergence after injection begins. \textbf{Plot3:} Distribution of the four failure modes across data formats and model scales. FP8 produces no Silent Degradation or Gradual Drift. Its outcomes are strictly binary (recover or crash). FP16 shows the highest Spike-and-Degrade fraction, consistent with its wider dynamic range enabling large-but-finite corruptions.}
    \label{fig:fault_loss_curves}
\end{figure*}

\subsection*{RQ1: How do permanent faults propagate through training, and what makes them different from transient faults?}
\label{sec:results:rq2}
% Show how a fault that goes to the output and affects PPL, goes to the output doesn't affect PPL, generates a NaN that goes to the output and generates a NaN that doesn't go to the output

%Prior work on faults during training injects a single transient bit-flip and observes whether training recovers or crashes~\cite{yu2025exploring,chen2025bit}. This answers whether training can survive a one-time event, but not what happens when a faulty GPU keeps corrupting computation over the course of training. We trace training under permanent fault injection and identify four distinct failure modes (Plot 1 in Figure~\ref{fig:fault_loss_curves}), only the first of which is observable under transient faults. 

Prior studies on transient faults during training typically inject a single bit flip and observe whether the training recovers or crashes~\cite{yu2025exploring,chen2025bit}. This probes resilience to a one-time perturbation, but not the behavior of a GPU that persistently corrupts computations over hundreds of thousands of iterations. Our pre-training runs are subjected to permanent fault injection, revealing four qualitatively distinct failure modes (Plot~1 in Figure~\ref{fig:fault_loss_curves}), only the first of which would be observable under a transient fault model.

%Each figure has two panels: the left shows the magnitude and type of injected bit flips per iteration (mantissa in blue, exponent in orange, sign bit in red), and the right shows the effect on training dynamics (LM loss, dynamic loss scale (DLS), or weight norm).

\textbf{1) Spike-and-recover.}
At low fault rates, corruptions are too infrequent to overwhelm the training dynamics. Each NaN event triggers mixed-precision recovery (loss-scale reduction and update skipping~\cite{micikevicius2017mixed}), after which gradient descent proceeds with predominantly clean updates. The loss therefore exhibits isolated spikes followed by return to the fault-free trajectory. This is the regime that single-injection studies capture and typically interpret as \emph{training is resilient} to the injected fault.

\textbf{2) Spike-and-degrade.}
At higher fault rates, NaN events become sufficiently frequent that the loss scale collapses and the skip-and-halve mechanism no longer restores stability. The optimizer is effectively stalled: many steps are skipped, the effective learning signal vanishes, and the loss rises monotonically as corrupted activations repeatedly propagate through the network. In contrast to spike-and-recover, fault rate here exceeds the threshold at which training dynamics can self-heal.

\textbf{3) Silent degradation.}
Faults are frequent and severe enough to corrupt the training state, but not enough to crash the job or consistently produce NaNs. Each faulty iteration perturbs activations or gradients while remaining finite, so weights are updated using biased signals and the loss drifts upward over time. The run completes without obvious instability indicators, making this behavior operationally difficult to detect.

\textbf{4) Gradual drift.}
Faults predominantly affect backward-pass quantities (e.g., gradient inputs), introducing small but systematic errors into each update step. These biases accumulate over long horizons, leading to a slow divergence from the fault-free trajectory without sharp loss spikes or NaNs. This mode is invisible to NaN-based monitoring and unlikely to be exposed by short fine-tuning runs, but becomes apparent over full pre-training durations.

\paragraph{Weight divergence reveals what loss curves hide:}
Plot~1 in Figure~\ref{fig:fault_loss_curves} shows the loss trajectories that a practitioner \emph{would} observe, whereas Plot~2 reports the L2 distance between the faulty and fault-free parameter vectors throughout training. All permanent fault runs exhibit monotonically increasing parameter divergence once injection begins; the model does not rejoin the baseline trajectory. Runs with error rates \(r \leq 0.005\) remain near-zero divergence, consistent with benign perplexity outcomes, while runs with \(r \geq 0.01\) show steadily growing distance. Notably, even trajectories whose loss remains near baseline (Gradual Drift) accumulate substantial weight drift, indicating that loss alone can mask significant deviations in the training parameters.

\paragraph{Failure mode prevalence across formats:}
Plot~3 summarizes how the four failure modes distribute across numerical formats and model scales. Two effects are most prominent:  FP8 produces no instances of Silent Degradation or Gradual Drift: outcomes are effectively binary (recover or crash), which is consistent with its narrow dynamic range forcing faults either to saturate harmlessly or to trigger NaNs; in contrast, FP16 exhibits the largest fraction of Spike-and-degrade runs - its wider dynamic range permits large but finite corruptions that degrade the model
without necessarily causing a crash. These format-dependent patterns indicate that numerical representation is a first-order determinant of how permanent faults manifest at the training-loss level.

\begin{figure}[ht!]
    \centering
    \includegraphics[width=\columnwidth]{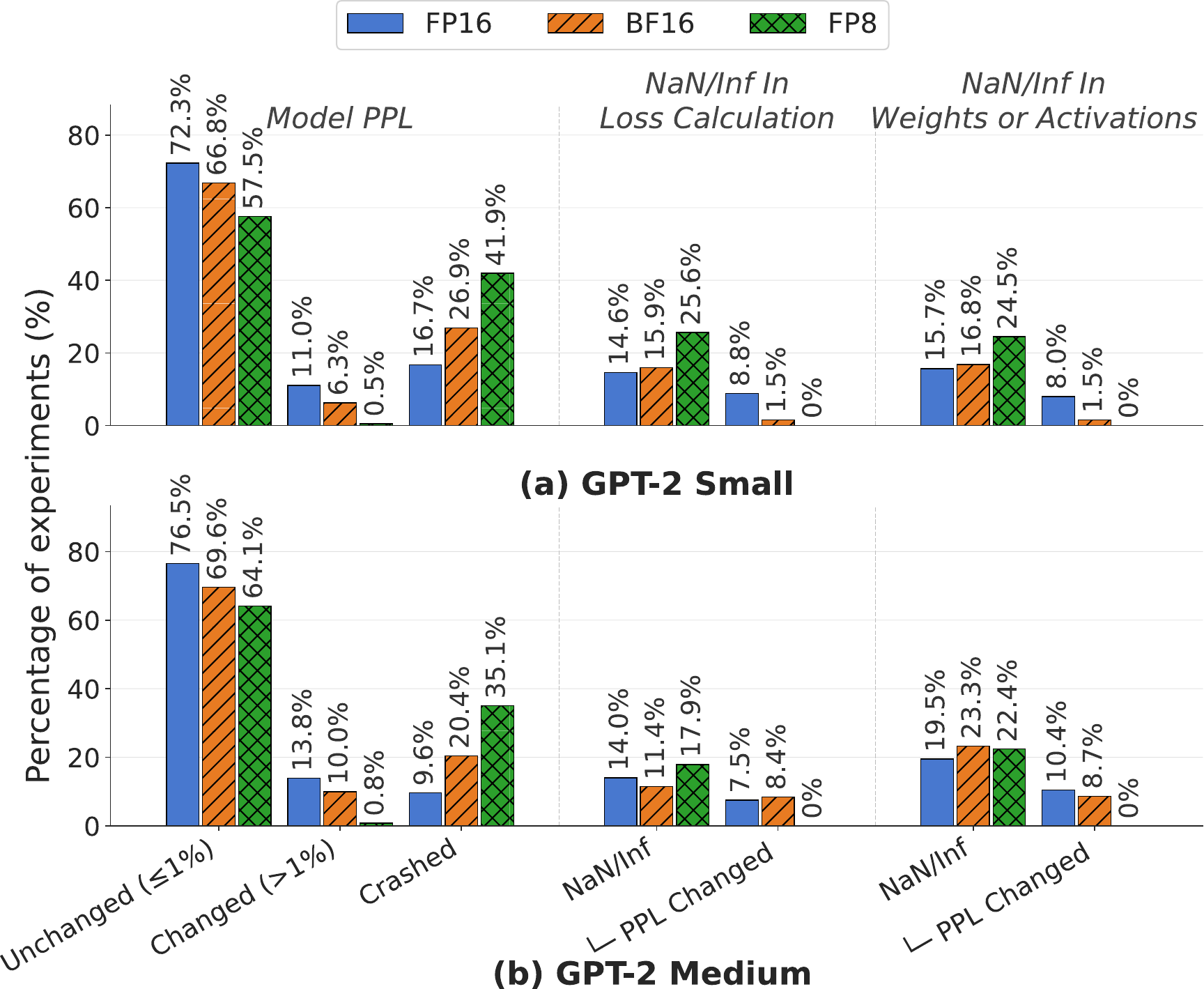}
    \caption{Distribution of training outcomes under permanent fault injection across data formats for GPT2-Small(a) and GPT2-Medium(b). Left: final model PPL classified as Unchanged, Changed, or Crashed. Middle and right: fraction of runs encountering NaN/Inf in loss or weights/activations, with sub-bars showing the subset that also resulted in a PPL change. Lower-precision formats crash more but degrade silently less; larger models show a higher fraction of silent degradation.}
    \label{fig:fault_dist}
\end{figure}

\subsection*{RQ2: How resilient is LLM training to permanent faults?}
%We study this question from three angles:

\paragraph{1) Model training dynamics under fault.}
We first characterize the impact of fault injection along three complementary dimensions summarized in Figure~\ref{fig:fault_dist}. The primary dimension is final model quality: we measure the perplexity (PPL) of the trained model and classify each run as \emph{Unchanged} (PPL within 1\% of the fault-free baseline of 28.54), \emph{Changed} (PPL deviates by more than 1\%), or \emph{Crashed} (training does not produce a valid PPL). The 1\% threshold is empirically motivated: repeated fault-free runs exhibit up to 1\% natural variance, so smaller deviations cannot be confidently attributed to faults.
The second dimension records whether the training loss becomes \textbf{NaN} or \textbf{Inf} at any iteration, indicating a breakdown of the optimization objective. The third dimension tracks NaN/Inf events in weights or activations, capturing instability propagating through the network independently of the loss. For the
latter two dimensions, a sub-bar reports what fraction of affected runs also exhibit a PPL change, linking low-level numerical faults to observable model degradation.

Across all formats, the dominant trend is that LLM training is remarkably resilient to permanent hardware faults: 57–76\% of runs complete with PPL indistinguishable from the fault-free baseline. Model behavior under failure, however, depends strongly on numerical precision. For GPT-2 Small, FP16’s remaining 28\% of runs split between graceful degradation (11.0\% PPL Changed) and crashes (16.7\%), with NaN/Inf loss events (14.6\%) serving as the primary propagation pathway: roughly 60\% of FP16 runs that encounter a NaN/Inf loss still finish training but with elevated PPL. 
BF16 and FP8 follow a different pattern: crash rates rise to 26.9\% and 41.9\%, while the fraction showing non-crashing PPL degradation falls to 6.3\% and 0.5\%. This inverse relationship between crash rate and PPL-change rate reflects a shift in failure mode: lower-precision formats have narrower dynamic ranges, so faults that
overflow to NaN typically produce catastrophic failure rather than recoverable, but degraded, solutions. Consequently, in BF16 and FP8,
NaN/Inf loss events almost always correspond to either a full crash or a near-baseline outcome, with little middle ground. The same trend appears for NaN/Inf in weights and activations, which closely track loss instability across formats, indicating that faults originate in the computational layers and propagate upward into the loss. GPT-2 Medium (Figure~\ref{fig:fault_dist}b) exhibits a similar qualitative pattern but with a larger Changed fraction (13.8\% for FP16 vs.\ 11.0\% for Small), suggesting that larger models provide more opportunities for faults to induce silent degradation rather than crashes.

\paragraph{2) Impact on PPL.}
We next examine which fault parameters drive variation in model quality. Figure~\ref{fig:ppl_vs} reports the distribution of final PPL for fault-injected runs, stratified by fault rate, injection checkpoint, and fault phase for FP16, BF16, and FP8; the fault-free baseline is PPL = 28.54. The following observations summarize the main trends:
\begin{enumerate}[label=\textbf{Obsv.~\arabic*:}, leftmargin=*, wide]
    \item \textbf{Most runs survive, but tail risk is severe.}
    Across all formats, the majority of runs converge near the baseline. For FP16 at fault rate 1.0, the median PPL is 28.58, yet the 95th percentile reaches 192.4, corresponding to converged but poor solutions rather than outright divergence. Permanent faults can therefore silently degrade models
    without triggering obvious failures.

    \item \textbf{Low fault rates are benign.}
    At rates 0.001 through 0.01, PPL is statistically indistinguishable from the baseline across all three formats. Degradation appears only at rates 0.05 and above, consistent with the optimizer absorbing infrequent corruptions through subsequent updates. Format-dependent differences in distributional spread become pronounced at higher rates.

    \item \textbf{Earlier faults cause more damage.}
    Faults injected at step 2571 yield broader PPL distributions than those at steps 5142 or 7713. Early in training, the model has not yet reached a stable region of the loss landscape and is therefore more sensitive to perturbations.

    \item \textbf{Backward weight gradients are resilient; forward outputs and backward input gradients are not.}
    Faults in \texttt{bwd\_grad\_wt} have little effect on PPL, due to two mitigating factors: optimizer momentum smooths corrupted gradients over multiple steps, and in distributed training, gradients are averaged across GPUs, diluting the impact of a single faulty rank by a factor of \(1/N\). In contrast, faults in \texttt{fwd\_out} and \texttt{bwd\_grad\_inp} act on local computation before any cross-rank communication and thus propagate without attenuation.
\end{enumerate}

\paragraph{3) Downstream performance.}
Perplexity is an imperfect proxy for task performance: models with similar PPL can differ substantially on downstream evaluations \cite{meister2021language}. To test whether training-time corruption from permanent faults translates into real capability loss, we evaluate GPT-2 Medium models using the LM Evaluation Harness~\cite{eval-harness} on the Children's Book Test (CBT)~\cite{hill2015goldilocks} and the Winograd Schema Challenge \cite{levesque2012winograd}. GPT-2 Small is excluded as it lacks sufficient capacity to produce meaningful scores on these tasks.

Figure~\ref{fig:down_perf} shows that downstream degradation is not confined to runs with obviously corrupted loss curves. 
%Using the failure-mode categories defined in RQ1, we evaluate representative models from each mode. As expected, 
The Spike-and-degrade models perform poorly across all benchmarks. The Spike-and-recover and Gradual Drift models, which maintain near-baseline PPL, still exhibit measurable accuracy drops on both tasks. These results confirm that permanent faults can harm learned representations in ways that are not visible in training perplexity alone, and that loss-curve monitoring is insufficient to detect all forms of training-time corruption induced by faulty hardware.

\begin{figure}
    \centering
    \includegraphics[width=0.8\columnwidth]{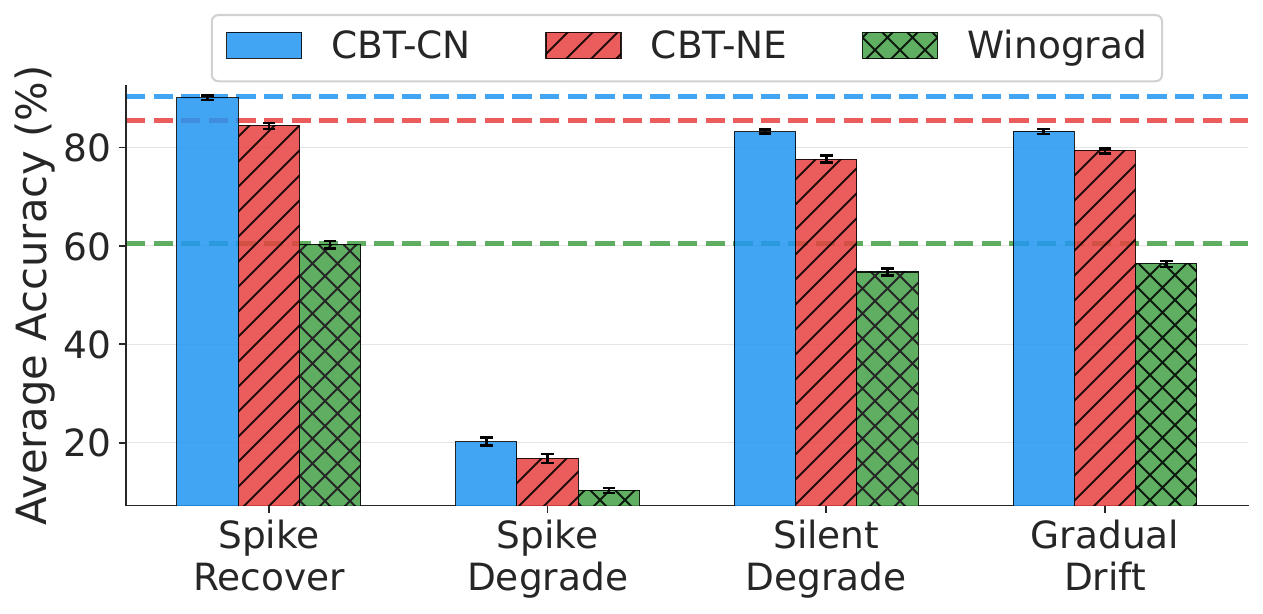}
    \caption{Performance of faulty GPT2-Medium (BF16) models on downstream tasks. We classify the models in the four categories and run them on Children's Book Test Dataset~\cite{hill2015goldilocks} (CN = Common Nouns and NE = Named Entities) and Winograd Schema Challenge~\cite{levesque2012winograd}. We can see that even for cases like spike recover or gradual drifts, the downstream performance is inferior to that of the fault free baselines on downstream tasks due to a change in the training dynamics.} 
    \label{fig:down_perf}
\end{figure}

\subsection*{RQ3: How does the choice of numerical data format influence the resilience of LLM training to permanent hardware faults?}
%Show the variation in various training outcomes with different format
\begin{figure*}[ht]
    \centering
    \includegraphics[width=1\linewidth]{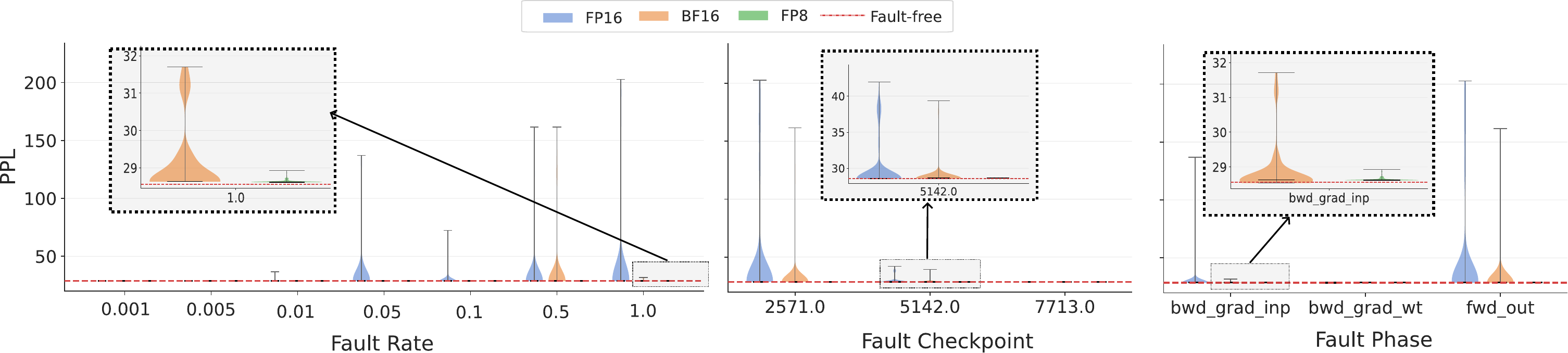}
    \caption{Variation in PPL studied against fault rate, fault checkpoint and fault phase. As the fault rate increases, we can see a variation in the PPL. Also, the earlier the fault is injected, the higher chances that it will manifest as a change in PPL. And, when faults are injected in the gradients wrt inputs or the forward pass, we see a higher change in PPL.}
    \label{fig:ppl_vs}
\end{figure*}

\paragraph{Data format strongly modulates resilience: FP8 $>$ BF16 $>$ FP16.}
Figure~\ref{fig:ppl_vs} shows a pronounced difference in PPL distributional spread across numerical formats. FP16 exhibits the widest distributions and most extreme outliers, especially at high fault rates. BF16 yields substantially tighter distributions under the same fault conditions, and FP8 is the most compact of the three. This ordering aligns with the structural properties of the formats: FP16 uses 1 sign, 5 exponent, and 10 mantissa bits, so a single exponent bit-flip can induce a large-magnitude change. BF16 retains 8 exponent bits but only 7 mantissa bits, matching the dynamic range of FP32 and reducing the worst-case impact of individual bit corruptions. FP8 further narrows the effective dynamic range, which bounds the magnitude of individual corruption events and provides a degree of fault containment at the cost of reduced numerical precision.

\subsection*{RQ4: How effective is loss-NaN checking in reducing the impact of permanent faults on training dynamics?}
% Show what happens to things like nan check to the output PPL

\begin{figure}
    \centering
    \includegraphics[width=0.95\columnwidth]{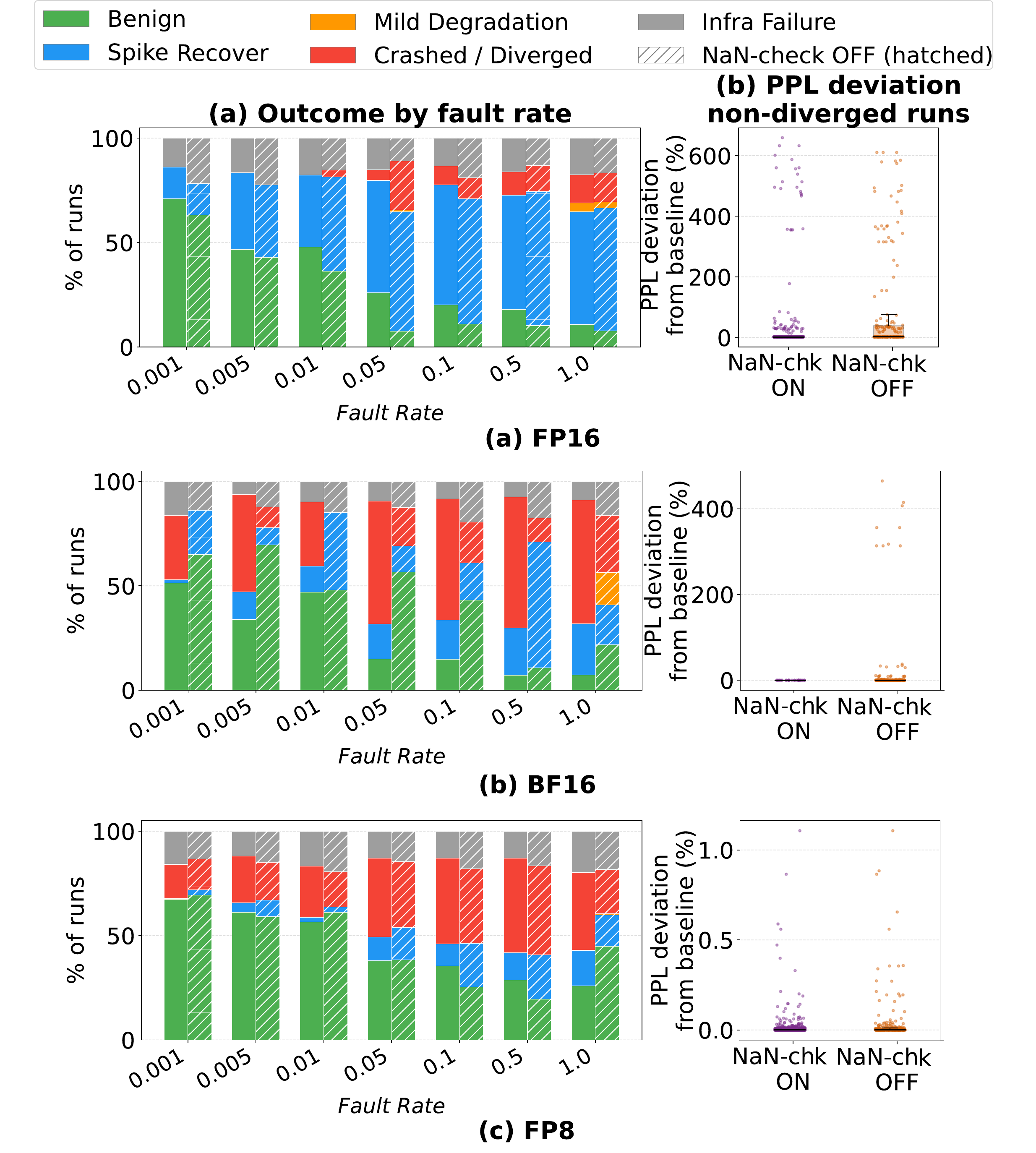}
    \caption{Effect of the loss NaN check across data formats. \textbf{Left:} outcome distribution by fault rate with NaN check enabled (solid) and disabled (hatched). \textbf{Right:} PPL deviation of non-diverged runs under both conditions. For FP16, the check converts many crashes into Spike-and-Recover events but extreme silent degradation persists. For BF16, the check mitigates both crashes and silent degradation. For FP8, outcomes are binary regardless of the check, with PPL deviations under 1\%.}
    \label{fig:rq4}
\end{figure}

The loss-NaN check is a standard mixed-precision safeguard: when a NaN or Inf is detected in the loss, the optimizer step is skipped and the dynamic loss scale is halved. We compare the training dynamics of all the three data formats under permanent faults and present our results in Figure~\ref{fig:rq4}.
For FP16, the failure spectrum is broad: Crashed/Diverged, Spike-and-recover, and Mild Degradation all appear across fault rates. With the NaN check enabled, Spike-and-recover (blue) becomes the dominant non-benign outcome, whereas without it Crashed/Diverged (red) dominates for rates \( \ge 0.05 \). This indicates that, for FP16, the check successfully intercepts a substantial fraction of training anomalies that would otherwise be fatal. 

%Second, fault phase modulates the benefit: Forward Output and Backward Gradient Input faults carry the highest crash risk under both conditions, while Backward Gradient Weight faults remain the most self-correcting, a pattern that holds across all three formats.

A particularly consequential effect is that non-diverged FP16 runs exhibit PPL deviations of up to \(\sim\)600\% above baseline (rightmost panel of Figure~\ref{fig:rq4}a), representing the largest silent degradation among all formats. The NaN check substantially tightens this distribution, but extreme outliers remain, indicating that some fault-induced corruptions accumulate below the NaN threshold and manifest as degraded model quality rather than as a detectable crash.

BF16 training (Figure~\ref{fig:rq4}b) exhibits a distinct risk profile that combines unfavorable aspects of both FP8 and FP16. Crash rates at high fault rates approach those of FP8, yet non-diverged runs can still suffer silent PPL degradation of up to \(\sim\)400\%, a failure mode largely absent in FP8. The NaN check mitigates both classes of failure: it reduces Crashed/Diverged outcomes and suppresses the large PPL deviations among surviving runs. This dual effect makes the check particularly important for BF16: disabling it increases outright failures and exposes training to a class of invisible quality loss that neither crashes nor triggers obvious alarms. As in FP16, the check mainly converts otherwise fatal trajectories to Spike-and-recover behavior rather than eliminating underlying faults.

\paragraph{Cross-format summary.}
FP8 fault outcomes are effectively binary: runs either absorb faults or diverge, so the NaN check primarily prevents crashes. FP16 training is most prone to persistent PPL degradation among surviving runs; the check reduces but does not eliminate this risk. BF16 combines high crash susceptibility with substantial silent degradation, both of which the check mitigates. Across formats and fault rates, fault phase is a stronger predictor of outcome than NaN checking: Forward Output and Backward Gradient Input faults consistently yield the highest crash rates (\(\sim\)60\%) at matched fault rates), whereas Backward Gradient Weight faults are largely self-correcting. This phase dependence appears format-agnostic, suggesting it is driven by gradient-flow structure rather than by the numerical representation.
\section{Limitations}
\label{sec:limit}

%- Lack of training on larger models
First, our evaluation is limited to relatively small-scale models, which may differ from the massive foundation models deployed in production clusters. We intentionally accept this trade-off: smaller models allow thousands of full training runs across diverse fault signatures, enabling a more rigorous characterization of SDC behavior.

%- still requires accurate hardware characterization
Second, the fidelity of our software-level fault injection depends on the accuracy of the underlying hardware defect models. Our conclusions are valid only when the simulated error signatures and activation rates reflect real silicon degradation; arbitrary bit-flip patterns would yield misleading results. Accurate error-signature characterization is therefore a prerequisite for meaningful use of the methodology.

Third, our experiments target a specific distributed training configuration with sequence parallelism and may not capture fault propagation under alternative parallelism strategies. Different pipeline and tensor-parallel layouts induce different communication and synchronization patterns, causing SDC on one node to propagate, attenuate, or amplify differently. Characterizing how distribution strategy shapes the SDC blast radius is left to future work.

\section{Conclusion}
\label{sec:conclusion}

%In this work, we propose a study setting that enables us to thoroughly investigate the effects of hardware faults on LLM training. \fixme{How do we enable this with our approach?}

%While prior work has studied fault resilience in deep learning primarily through application-level transient fault injection, such analyses are fundamentally constrained by idealized fault models based on arbitrary bit flips and simplified soft-error assumptions that diverge from the error signatures produced by real degraded hardware. In parallel, extensive device- and circuit-level studies have characterized aging mechanisms, early-life failures, and manufacturing defect physics, yet these results have remained largely decoupled from the application-level behavior that ultimately matters for modern AI training workloads. This work bridges that gap by linking physically grounded defect and aging mechanisms directly to their observable impact on end-to-end LLM pre-training dynamics.

This paper presents the first systematic characterization of large language model training sensitivity to SDC caused by permanent and intermittently manifesting hardware faults.
Our results show that, although SDC-induced perturbations to submodule computations and gradients are often numerically small, they can still steer optimization toward different local optima and yield models with meaningfully different learned weights and downstream behavior. We further demonstrate that SDC can be highly evasive under conventional monitoring: in many cases, training loss appears well behaved, masking underlying corruption, while in extreme cases, the same class of faults can trigger sharp loss spikes and catastrophic divergence. We find that SDC impact is strongly dependent on the specific unhealthy node and fault site, and is tightly coupled to the local loss landscape of the training task, leading to highly heterogeneous resilience behavior even within a single cluster.
Therefore, this work establishes an empirical foundation for reasoning about detection strategies, resilience targets, and system-level reliability requirements for LLM training at scale.
Beyond the vulnerability assessment of the training loop, this work offers a methodology that future work can leverage towards integrating hardware-accurate fault models and realistic intermittent activation behavior with software-level training analysis to co-design hardware, runtime, and optimization algorithms for fault-aware large-scale AI systems.

\bibliographystyle{ieeetr}
\bibliography{IEEEexample}

\end{document}